\documentclass{llncs}
\usepackage{amsmath,amssymb,amsfonts}
\usepackage{mathtools}
\usepackage{thm-restate}
\usepackage{adjustbox}      
\usepackage[usenames,dvipsnames]{color}
\usepackage{xcolor}
\usepackage{xspace}
\usepackage{microtype}
\usepackage{todonotes}
\usepackage{colonequals}
\usepackage{wasysym}
 \usepackage{booktabs}

\usepackage{paralist}
\usepackage{multirow}
\usepackage{multicol}

\usepackage{refcount}
\usepackage{paralist}

\usepackage{subfigure}
\usepackage{url}
\usepackage{appendix}
\usepackage{graphicx}
\usepackage{float}

\usepackage[vlined,linesnumbered]{algorithm2e}
\SetAlCapSkip{2ex plus 0.5ex minus 1ex}
\SetEndCharOfAlgoLine{}
\SetArgSty{textrm}
\SetKwInOut{Input}{Input}\SetKwInOut{Output}{Output}
\SetCommentSty{textit}
\SetKw{Break}{break}
\SetKwProg{Type}{type}{}{end}
\SetKwProg{Function}{function}{}{end}
\SetKwFor{ForEach}{foreach}{do}{}
\SetKwFor{WhileTrue}{repeat}{}{end}
\SetKw{Continue}{continue}
\SetKwRepeat{DoWhile}{repeat}{while}

\usepackage{listings}
\usepackage{nicefrac}
\usepackage{tikz} 
\usepackage{pgfplots, pgfplotstable}
\usepackage{marvosym}
\usepackage{wrapfig}
\usepackage{hyperref}

\usepackage{footnote}
\usepackage{tablefootnote}
\usepackage{array}

\usetikzlibrary{arrows,decorations.pathmorphing,positioning,fit,trees,shapes,shadows,automata,calc,patterns}

\usepackage[capitalise]{cleveref}

\input{macros} 

\newtoggle{tr}
\toggletrue{tr}

\begin{document}
	

  \title{Sound Value Iteration%
\thanks{This work is partially supported by the
		Sino-German Center project CAP (GZ 1023).}}

\author{Tim Quatmann, Joost-Pieter Katoen}

 \institute{RWTH Aachen University, Aachen, Germany}

\maketitle

  \begin{abstract}
Computing reachability probabilities is at the heart of probabilistic
model checking. All model checkers compute these probabilities in an
iterative fashion using value iteration. This technique approximates
a fixed point from below by determining reachability probabilities for
an increasing number of steps. To avoid results that are significantly
off, variants have recently been proposed that converge from both below
and above. These procedures require starting values for both sides. We
present an alternative that does not require the a priori computation of
starting vectors and that converges faster on many benchmarks. The crux
of our technique is to give tight and safe bounds --- whose computation
is cheap --- on the reachability probabilities. Lifting this technique
to expected rewards is trivial for both Markov chains and MDPs.
Experimental results on a large set of benchmarks show its scalability
and efficiency.
\end{abstract}

  \section{Introduction}

Markov decision processes (MDPs)~\cite{Put94,mdp_handbook2002} have their roots in operations research and stochastic control theory. 
They are frequently used for stochastic and dynamic optimization problems and are widely applicable in, e.g., stochastic scheduling and robotics.
MDPs are also a natural model in randomized distributed computing where coin flips by the individual processes are mixed with non-determinism arising from interleaving the processes' behaviors.
The central problem for MDPs is to find a policy that determines what action to take in the light of what is known about the system at the time of choice. 
The typical aim is to optimize a given objective, such as minimizing the expected cost until a given number of repairs, maximizing the probability of being operational for 1,000 steps, or minimizing the probability to reach a ``bad'' state.

Probabilistic model checking~\cite{DBLP:conf/lics/Katoen16,DBLP:series/natosec/Baier16} provides a scalable alternative to tackle these MDP problems, see the recent surveys~\cite{Etessami2016,DBLP:book/BdAFK18}.
The central computational issue in MDP model checking is to solve a system of linear inequalities. 
In absence of non-determinism --- the MDP being a Markov Chain (MC) --- a linear equation system is obtained.
After appropriate pre-computations, such as determining the states for which no policy exists that eventually reaches the goal state, the (in)equation system has a unique solution that coincides with the extremal value that is sought for.
Possible solution techniques to compute such solutions include policy iteration, linear programming, and value iteration.
Modern probabilistic model checkers such as \prism~\cite{DBLP:conf/cav/KwiatkowskaNP11} and \storm~\cite{DBLP:conf/cav/DehnertJK017} use value iteration by default.
This approximates a fixed point from below by determining the probabilities to reach a target state within $k$ steps in the $k$-th iteration.
The iteration is typically stopped if the difference between the value vectors of two successive (or vectors that are further apart) is below the desired accuracy $\precision$.

This procedure however can provide results that are significantly off, as the iteration is stopped prematurely, e.g., since the probability mass in the MDP only changes slightly in a series of computational steps due to a ``slow'' movement.
This problem is not new; similar problems, e.g., occur in iterative approaches to compute long-run averages~\cite{DBLP:conf/qest/KatoenZ06} and transient measures~\cite{DBLP:journals/pe/Malhotra96} and pop up in statistical model checking to decide when to stop simulating for unbounded reachability properties~\cite{DBLP:journals/tocl/DacaHKP17}. 
As recently was shown, this phenomenon does not only occur for hypothetical cases but affects practical benchmarks of MDP model checking too~\cite{DBLP:conf/cav/Baier0L0W17}.
To remedy this, Haddad and Monmege~\cite{HADDAD2017} proposed to iteratively approximate the (unique) fixed point from both below and above; a natural termination criterion is to halt the computation once the two approximations differ less than $2{\cdot}\precision$.
This scheme requires two starting vectors, one for each approximation. 
For reachability probabilities, the conservative values zero and one can be used.
For expected rewards, it is non-trivial to find an appropriate upper bound --- how to ``guess'' an adequate upper bound to the expected reward to reach a goal state?
Baier \emph{et al.}~\cite{DBLP:conf/cav/Baier0L0W17} recently provided an algorithm to solve this issue.

This paper takes an alternative perspective to obtaining a sound variant of value iteration.
\emph{Our approach does not require the a priori computation of starting vectors and converges faster on many benchmarks.} 
The crux of our technique is to give tight and safe bounds --- whose computation is cheap and that are obtained during the course of value iteration --- on the reachability probabilities.
The approach is simple and can be lifted straightforwardly to expected rewards.
The central idea is to split the desired probability for reaching a target state into the sum of
\begin{enumerate}[(i)]
	\item\label{enum:intro:withink} the probability for reaching a target state \emph{within} $k$ steps and
	\item\label{enum:intro:afterk} the probability for reaching a target state \emph{only after} $k$ steps.
\end{enumerate}
We obtain~(\ref{enum:intro:withink}) via $k$ iterations of (standard) value iteration.
A second instance of value iteration computes the probability that a target state is still reachable after $k$ steps.
We show that from this information safe lower and upper bounds for (\ref{enum:intro:afterk}) can be derived.
%
%
%
We illustrate that the same idea can be applied to expected rewards, topological value iteration~\cite{DBLP:journals/jair/DaiMWG11}, and Gauss-Seidel value iteration.
We also discuss in detail its extension to MDPs and provide extensive experimental evaluation using our implementation in the model checker \storm~\cite{DBLP:conf/cav/DehnertJK017}.
Our experiments show that on many practical benchmarks we need significantly fewer iterations, yielding a speed-up of about 20\% on average.
More importantly though, is the conceptual simplicity of our approach.

  \section{Preliminaries}
\label{sec:prelim}

	For a finite set $\States$ and vector $x \in \RR^{|\States|}$, let $\vectoraccess{x}{s} \in \RR$ denote the entry of $x$ that corresponds to $s \in \States$.
	Let $\States' \subseteq \States$ and $a \in \RR$.
	We write $\vectoraccessSet{x}{\States'} = a$ to denote that $\vectoraccess{x}{s} = a$ for all $s \in \States'$.
	Given $x,y \in \RR^{|\States|}$, $x \le y$ holds iff $\vectoraccess{x}{s} \le \vectoraccess{y}{s}$ holds for all $s \in \States$.
	For a function $f \colon \RR^{|\States|} \to \RR^{|\States|}$ and $k \ge 0$ we write  $f^k$ for the function obtained by applying $f$ $k$ times, i.e., $f^0(x) = x$ and $f^k(x) = f(f^{k-1}(x))$ if $k>0$. 

\subsection{Probabilistic Models and Measures}
\label{sec:prelim:modelsmeasures}
We briefly present probabilistic models and their properties. More details can be found in, e.g.,~\cite{BK08}.
\begin{definition}[Probabilistic Models]
	A \emph{Markov Decision Process (MDP)} is a tuple $\mdpDef$, where
	\begin{compactitem}
		\item $\States$ is a finite set of states, $\Actions$ is a finite set of actions, $\sinit$ is the initial state, 
		\item $\probP \colon \States \times \Actions \times \States \to [0,1]$ is a transition probability function satisfying\linebreak$\sum_{s' \in \States} \probP(s, \act, s') \in \{0,1\} $ for all $s \in \States, \act \in \Actions$, and
		\item $\rewFct \colon \States \times \Actions \to \RR$ is a reward function.
	\end{compactitem}
	$\mdp$ is a \emph{Markov Chain (MC)} if $|\Actions| = 1$.
\end{definition}
	
\begin{example}
 \cref{fig:prelim:models} shows an example MC and an example MDP.
\end{example}
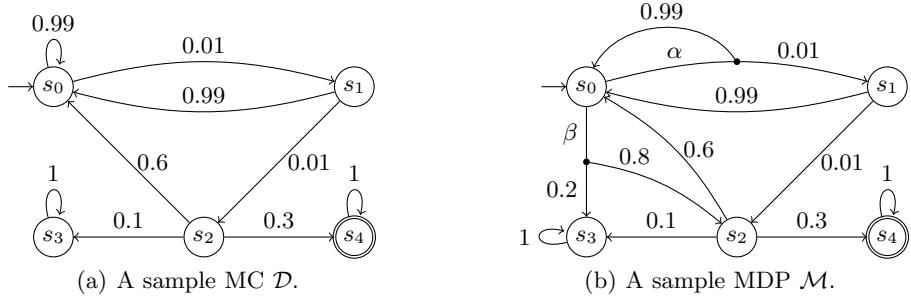
\begin{figure}[t]
	\centering
	\subfigure[A sample MC $\dtmc$.]{
		\label{fig:prelim:dtmc}
		\begin{tikzpicture}[scale=1]
		\node [state] (s0) at (0,0) {$s_0$};
		\node [state] (s1) [on grid, right =40mm of s0] {$s_1$};
		\node [state] (s3) [on grid, below=20mm of s0] {$s_3$};
		\node [state] (s2) [on grid, right=20mm of s3] {$s_2$};
		\node [state, target] (s4) [on grid, right=20mm of s2] {$s_4$};
		
		\initstateLeft{s0}
		
		\draw (s0) edge[transition, bend left=15] node [above] { 0.01} (s1);
		\draw (s0) edge [transition, loop above] node[above] {0.99} (s0);
		\draw (s1) edge[transition] node[right] {0.01} (s2);
		\draw (s1) edge[transition,bend left=15] node[above] {0.99} (s0);
		\draw (s2) edge[transition] node[above] {0.3} (s4);
		\draw (s2) edge[transition] node[above] {0.1} (s3);
		\draw (s2) edge[transition] node[right] {0.6} (s0);
		\draw (s3) edge [transition, loop above] node[above] {1} (s3);
		\draw (s4) edge [transition, loop above] node[above] {1} (s4);
		\end{tikzpicture}}%
	\hfill
	\subfigure[A sample MDP $\mdp$.]{
		\label{fig:prelim:mdp}
		\begin{tikzpicture}[scale=1]
		\node [state] (s0) at (0,0) {$s_0$};
		\node [state] (s1) [on grid, right =40mm of s0] {$s_1$};
		\node [state] (s3) [on grid, below=20mm of s0] {$s_3$};
		\node [state] (s2) [on grid, right=20mm of s3] {$s_2$};
		\node [state, target] (s4) [on grid, right=20mm of s2] {$s_4$};
		\initstateLeft{s0}
		
		\draw (s1) edge[transition] node[right] {0.01} (s2);
		\draw (s1) edge[transition,bend left=15] node[above] {0.99} (s0);
		\draw (s2) edge[transition] node[above] {0.3} (s4);
		\draw (s2) edge[transition] node[above] {0.1} (s3);
		\draw (s2) edge[transition, bend right=15] node[right] {0.6} (s0);
		\draw (s3) edge [transition, loop left] node[left] {1} (s3);
		\draw (s4) edge [transition, loop above] node[above] {1} (s4);
		\draw (s0) edge[transition, bend left=15] node[pos=0.25, above] {$\act$} node[action] (s0alpha) {} node [pos=0.75, above] { 0.01} (s1);
		\draw (s0alpha) edge[transition, bend right=60] node[above] {0.99} (s0);
		\draw (s0) edge[transition] node[pos=0.25, left] {$\altact$} node[action] (s0beta) {} node [pos=0.75, left] { 0.2} (s3);
		\draw (s0beta) edge[transition, bend left=15] node[above, pos=0.3] {0.8} (s2);
		\end{tikzpicture}
	}
	\caption{Example models.}
		\label{fig:prelim:models}
\end{figure}
	We often simplify notations for MCs by omitting the (unique) action.
	For an MDP $\mdpDef$, the set of \emph{enabled actions} of state $s \in \States$ is given by $\Act{s} = \{ \act \in \Actions \mid \sum_{s' \in \States} \probP(s, \act, s') = 1  \}$.
	We assume that $\Act{s} \neq \emptyset$ for each $s \in \States$.
	Intuitively, upon performing action $\act$ at state $s$ reward $\rewFct(s,\act)$ is collected and with probability $\probP(s, \act, s')$ we move to $s' \in \States$.
	Notice that rewards can be positive or negative.
	
	A state $s \in \States$ is called \emph{absorbing} if  $\probP(s,\act,s) = 1$ for every $\act \in \Act{s}$.
	A \emph{path} of $\mdp$ is an infinite alternating sequence $\ppath = s_0 \act_0 s_1 \act_1 \dots$ where $s_i \in \States$, $\act_i \in \Act{s_i}$, and $\probP(s_i, \act_i, s_{i+1}) > 0$ for all $i\ge 0$.
	The set of paths of $\mdp$ is denoted by $\Paths[\mdp]$.
	The set of paths that start at $s \in \States$ is given by $\Paths[\mdp,s]$.
	A \emph{finite path} $\ppathfin = s_0 \act_0 \dots \act_{n-1} s_n$ is a finite prefix of a path ending with $\last{\ppathfin} = s_n \in \States$. $|\ppathfin| = n$ is the length of $\ppathfin$, $\Pathsfin[\mdp]$ is the set of finite paths of $\mdp$, and $\Pathsfin[\mdp,s]$ is the set of finite paths that start at state $s\in\States$.
	We consider LTL-like notations for sets of paths.
	For $k \in \NN \cup \{\infty\}$ and $G, H \subseteq \States$ let
	\[
	H \until^{\le k} G = \{ s_0 \act_0 s_1 \dots  \in \Paths[\mdp,\sinit] \mid s_0, \dots, s_{j-1} \in H \text{, } s_j \in G \text{ for some } j \le k\}
	\]
	denote the set of paths that, starting from the initial state $\sinit$, only visit states in $H$ until after at most $k$ steps a state in $G$ is reached. Sets $H \until^{> k} G$ and $H \until^{=k} G$ are defined similarly.
	We use the shorthands $\eventually^{\le k} G \coloneqq \States \until^{\le k} G$, $\eventually G \coloneqq \eventually^{\le \infty} G$, and $\globally^{\le k} G \coloneqq \Paths[\mdp, \sinit] \setminus \eventually^{\le k} (\States \setminus G)$. 
	
	A \emph{(deterministic) scheduler} for $\mdp$ is a function $\sched \colon \Pathsfin[\mdp] \to \Actions$ such that
	$\sched(\ppathfin) \in \Act{\last{\ppathfin}}$ for all $\ppathfin \in \Pathsfin[\mdp]$.
	The set of (deterministic) schedulers for $\mdp$ is $\Scheds[\mdp]$.
	$\sched \in \Scheds[\mdp]$ is called \emph{positional} if $\sched(\ppathfin)$ only depends on the last state of $\ppathfin$, i.e., for all $\ppathfin, \ppathfin' \in \Pathsfin[\mdp]$ we have $\last{\ppathfin} = \last{\ppathfin'}$ implies $\sched(\ppathfin) = \sched(\ppathfin')$.
	For MDP $\mdp$ and scheduler $\sched \in \Scheds[\mdp]$ the \emph{probability measure}   over finite paths is given by $\Prob_\mathit{fin}^{\mdp, \sched} \colon \Pathsfin[\mdp, \sinit] \to [0,1]$ with 
	$
	\Prob_\mathit{fin}^{\mdp, \sched} (s_0 \dots s_n) = \prod_{i=0}^{n-1} \probP(s_i, \sched(s_0\dots s_i),  s_{i+1}).
	$ 
	The probability measure $\Prob^{\mdp, \sched}$ over measurable sets of infinite paths is obtained via a standard cylinder set construction~\cite{BK08}.
	\begin{definition}[Reachability Probability]
	The \emph{reachability probability} of MDP $\mdpDef$, $G \subseteq S$, and $\sched \in \Scheds[\mdp]$ is given by $\Prob^{\mdp, \sched}(\eventually G)$.
	\end{definition}
	For $k \in \NN \cup \{\infty\}$, the function $\eventuallyrew^{\le k} G \colon \eventually G \to \RR$ yields the $k$-bounded reachability reward of a path $\ppath = s_0 \act_0 s_1 \dots  \in \eventually G$.  
	We set $\eventuallyrew^{\le k} G(\ppath) = \sum_{i = 0}^{j-1} \rewFct(s_i, \act_i)$, where $j = \min (\{i\ge 0 \mid s_i \in G  \} \cup \{k\})$.
	We write $\eventuallyrew G$ instead of $\eventuallyrew^{ \le \infty} G$.
	\begin{definition}[Expected Reward]
	The \emph{expected (reachability) reward} of MDP $\mdpDef$, $G \subseteq S$, and $\sched \in \Scheds[\mdp]$ with $\Prob^{\mdp, \sched}(\eventually G) = 1$ is given by the expectation $\Exp^{\mdp, \sched}(\eventuallyrew G) = \int_{\ppath \in \eventually G} \eventuallyrew G (\ppath) \diff\Prob^{\mdp, \sched}(\ppath)$.
	\end{definition}
	We write $\Prob^{\mdp, \sched}_s$ and $\Exp^{\mdp, \sched}_s$ for the probability measure and expectation obtained by changing the initial state of $\mdp$ to $s \in \States$.
	If $\mdp$ is a Markov chain, there is only a single scheduler. In this case we may omit the superscript $\sched$ from $\Prob^{\mdp, \sched}$ and $\Exp^{\mdp, \sched}$.
	We also omit the superscript $\mdp$ if it is clear from the context.
	The maximal reachability probability of $\mdp$ and $G$ is given by $\Prob^{\mathrm{max}}(\eventually G) =  \max_{\sched\in\Scheds} \Prob^{\sched}(\eventually G)$.
	There is a a positional scheduler that attains this maximum~\cite{DBLP:journals/mor/BertsekasT91}.
	The same holds for minimal reachability probabilities and maximal or minimal expected rewards.

\begin{example}
	\label{ex:prelim:property}
	Consider the MDP $\mdp$ from \cref{fig:prelim:mdp}.
	We are interested in the maximal probability to reach state $s_4$ given by $\ProbSched[\mathrm{max}](\eventually \{s_4 \})$.
	Since $s_4$ is not reachable from $s_3$ we have $\ProbSched[\mathrm{max}]_{s_3}(\eventually \{s_4\}) = 0$.
	Intuitively, choosing action $\altact$ at state $s_0$ makes reaching $s_3$ more likely, which should be avoided in order to maximize the probability to reach $s_4$.
	We therefore assume a scheduler $\sched$ that always chooses action $\act$ at state $s_0$.
	Starting from the initial state $s_0$, we then eventually take the transition from $s_2$ to $s_3$ or the transition from $s_2$ to $s_4$  with probability one.
	The resulting probability to reach $s_4$ is given by $\ProbSched[\mathrm{max}](\eventually \{s_4 \}) = \ProbSched(\eventually \{s_4 \}) = 0.3/ (0.1 + 0.3) = 0.75$. 
\end{example}
	
	\subsection{Probabilistic Model Checking via Interval Iteration}
	\label{sec:prelim:modelcheck}
	
	In the following we present approaches to compute reachability probabilities and expected rewards.
	We consider approximative computations. Exact computations are handled in e.g.~\cite{DBLP:journals/corr/abs-1207-1264,DBLP:conf/fmcad/BauerMCS017}
	For the sake of clarity, we focus on reachability probabilities and sketch how the techniques can be lifted to expected rewards.
	
	\subsubsection{Reachability Probabilities.}
	We fix an MDP $\mdpDef$, a set of goal states $G \subseteq \States$, and a precision parameter $\precision > 0$.
	\begin{problem}\label{problem:reachprob}
		Compute an $\precision$-approximation of the maximal reachability probability $\reachprSched[\mathrm{max}]$, i.e., compute a value $\result \in [0,1]$ with  $|\result - \reachprSched[\mathrm{max}] | < \precision$.
	\end{problem}
	We briefly sketch how to compute such a value $\result$ via \emph{interval iteration} \cite{DBLP:conf/cav/Baier0L0W17,HADDAD2017,DBLP:conf/atva/BrazdilCCFKKPU14}.
	The computation for minimal reachability probabilities is analogous.
	
	W.l.o.g. it is assumed that the states in $G$ are absorbing.
	Using graph algorithms, we compute $\StatesZero = \{ s \in \States \mid \Prob^\mathrm{max}_s(\eventually G)  =0 \}$ and partition the state space of $\mdp$ into $\States = \StatesZero \cupdot G \cupdot \StatesMaybe$ with $\StatesMaybe = \States \setminus (G \cup \StatesZero)$.
	If   $\sinit \in \StatesZero$ or $\sinit \in G$, the probability $\Prob^\mathrm{max}(\eventually G)$ is 0 or 1, respectively. 
	From now on we assume $\sinit \in \StatesMaybe$.
	
	We say that $\mdp$ is \emph{contracting} with respect to $\States' \subseteq \States$ if $\Prob_s^{\sched}(\eventually \States') = 1$ for all $s \in \States$ and for all $\sched \in \Scheds[\mdp]$.
	We assume that $\mdp$ is contracting with respect to $G \cup \StatesZero$.
	Otherwise, we apply a transformation on the so-called \emph{end components}\footnote{Intuitively, an end component is a set of states $\States' \subseteq \States$ such that there is a scheduler inducing that from any $s \in \States'$ exactly the states in $\States'$ are visited infinitely often.} of $\mdp$, yielding a contracting MDP $\mdp'$ with the same maximal reachability probability as $\mdp$.
	Roughly, this transformation replaces each end component of $\mdp$ with a single state whose enabled actions coincide with the actions that previously lead outside of the end component.
	This step is detailed in~\cite{HADDAD2017,DBLP:conf/atva/BrazdilCCFKKPU14}.
	
	We have $\vectoraccess{x^*}{s} = \Prob^\mathrm{max}_s(\eventually G)$ for $s \in \States$ and the unique fixpoint $x^*$ of the function $f \colon \RR^{|\States|} \to \RR^{|\States|}$ with 
	$\vectoraccessSet{f(x)}{\StatesZero} = 0$, $\vectoraccessSet{f(x)}{G} = 1$, and 
	\[
	\vectoraccess{f(x)}{s} =
	\max_{\act \in \Act{s}} \sum_{s' \in \States} \probP(s, \act, s') \cdot \vectoraccess{x}{s'} 
	\]
	for $s \in \StatesMaybe$.
	Hence, computing $\reachprSched[\mathrm{max}]$ reduces to finding the fixpoint of $f$.
	
	A popular technique for this purpose is the \emph{value iteration} algorithm~\cite{Put94}.
	Given a starting vector $x \in \RR^{|\States|}$ with  $\vectoraccessSet{x}{\StatesZero} = 0$ and $\vectoraccessSet{x}{G} = 1$, standard value iteration computes $f^k(x)$ for increasing $k$ until $\max_{s\in\States} |\vectoraccess{f^k(x)}{s} - \vectoraccess{f^{k-1}(x)}{s}| < \varepsilon$ holds for a predefined precision $\varepsilon > 0$.
	As pointed out in, e.g.,~\cite{HADDAD2017}, there is no guarantee on the preciseness of the result $\result = \vectoraccess{f^k(x)}{\sinit}$, i.e., standard value iteration does not give any evidence on the error $|\result - \reachprSched[\mathrm{max}]|$.
	The intuitive reason is that value iteration only approximates the fixpoint $x^*$ from one side, yielding no indication on the distance between the current result and $x^*$.
\begin{example}
	Consider the MDP $\mdp$ from \cref{fig:prelim:mdp}.
	We invoked standard value iteration in \prism~\cite{DBLP:conf/cav/KwiatkowskaNP11} and \storm~\cite{DBLP:conf/cav/DehnertJK017} to compute the reachability probability $\ProbSched[\mathrm{max}](\eventually \{s_4 \})$.
	Recall from \cref{ex:prelim:property} that the correct solution is $0.75$.
	With (absolute) precision $\precision = 10^{-6}$ both model checkers returned $0.7248$.
	Notice that the user can improve the precision by considering, e.g., $\precision = 10^{-8}$ which yields $0.7497$. However, there is no guarantee on the preciseness of a given result.
\end{example}
	
	The \emph{interval iteration} algorithm~\cite{HADDAD2017,DBLP:conf/cav/Baier0L0W17,DBLP:conf/atva/BrazdilCCFKKPU14} addresses the impreciseness of value iteration.
	The idea is to approach the fixpoint $x^*$ from below and from above.
	The first step is to find starting vectors $x_\ell, x_u \in \RR^{|\States|}$ satisfying $\vectoraccessSet{x_\ell}{\StatesZero} = \vectoraccessSet{x_u}{\StatesZero} = 0$, $\vectoraccessSet{x_\ell}{G} = \vectoraccessSet{x_u}{G} = 1$, and $x_\ell \le x^* \le x_u$.
	As the entries of $x^*$ are probabilities, it is always valid to set $\vectoraccessSet{x_\ell}{\StatesMaybe} = 0$ and $\vectoraccessSet{x_u}{\StatesMaybe} = 1$.
	We have $f^k(x_\ell) \le x^* \le f^k(x_u)$ for any $k \ge 0$.
	Interval iteration computes $f^k(x_\ell)$ and $f^k(x_u)$ for increasing $k$ until $\max_{s\in\States} |\vectoraccess{f^k(x_\ell)}{s} - \vectoraccess{f^{k}(x_u)}{s}| < 2 \varepsilon$.
	For the result $\result = \nicefrac{1}{2} \cdot (\vectoraccess{f^k(x_\ell)}{\sinit} + \vectoraccess{f^k(x_u)}{\sinit}) $ we obtain that $|\result- \reachprSched[\mathrm{max}]| < \varepsilon$, i.e., we get a sound approximation of the maximal reachability probability.
	
	\begin{example}
We invoked interval iteration in \prism and \storm to compute the reachability probability $\ProbSched[\mathrm{max}](\eventually \{s_4 \})$ for the MDP $\mdp$ from \cref{fig:prelim:mdp}.
Both implementations correctly yield an $\precision$-approximation of $\ProbSched[\mathrm{max}](\eventually \{s_4 \})$, where we considered $\precision = 10^{-6}$.
However, both \prism and \storm required roughly 300,000 iterations for convergence.
	\end{example}
	
	\subsubsection{Expected Rewards.}
	Whereas \cite{HADDAD2017,DBLP:conf/atva/BrazdilCCFKKPU14} only consider reachability probabilities, \cite{DBLP:conf/cav/Baier0L0W17} extends interval iteration to  compute expected rewards.
	Let $\mdp$ be an MDP and $G$ be a set of absorbing states 
	such that $\mdp$ is contracting with respect to $G$.
	\begin{problem}\label{problem:exprew}
		Compute an $\precision$-approximation of the maximal expected reachability reward $\exprewSched[\mathrm{max}]$, i.e., compute a value $\result \in \RR$ with  $|\result - \exprewSched[\mathrm{max}] | < \precision$.
	\end{problem}
	We have $\vectoraccess{x^*}{s} = \Exp_s^\mathrm{max}(\eventuallyrew G)$ for the unique fixpoint $x^*$ of $g \colon \RR^{|\States|} \to \RR^{|\States|}$ with 
	\[
	\vectoraccessSet{g(x)}{G} = 0 \ \text{ and } \ 
	\vectoraccess{g(x)}{s} =
	\max_{\act \in \Act{s}}  \rewFct(s,\act) + \sum_{s' \in \States} \probP(s, \act, s') \cdot \vectoraccess{x}{s'} 
	\]
	for $s \notin G$.
	As for reachability probabilities, interval iteration can be applied to approximate this fixpoint.
	The crux lies in finding appropriate starting vectors $x_\ell, x_u \in \RR^{|\States|}$ guaranteeing  $x_\ell \le x^* \le x_u$.
	To this end, \cite{DBLP:conf/cav/Baier0L0W17} describe graph based algorithms that give an upper bound on the expected number of times each individual state $s \in \States \setminus G$ is visited.
	This then yields an approximation of the expected amount of reward collected at the various states.
  \section{Sound Value Iteration for MCs}
\label{sec:dtmc}
We present an algorithm for computing reachability probabilities and expected rewards as in Problems \ref{problem:reachprob} and \ref{problem:exprew}.
The algorithm is an alternative to the interval iteration approach~\cite{HaddadM14,DBLP:conf/cav/Baier0L0W17} but
(i) does not require an a priori computation of starting vectors $x_\ell, x_u \in \RR^{|S|}$ and 
(ii) converges faster on many practical benchmarks as shown in \cref{sec:experiments}.
For the sake of simplicity, we first restrict to computing reachability probabilities on MCs.

In the following, let $\dtmcDef$ be an MC, $G \subseteq \States$ be a set of absorbing goal states and $\precision > 0$ be a precision parameter.
We consider the partition $\States = \StatesZero \cupdot G \cupdot \StatesMaybe$ as in \cref{sec:prelim:modelcheck}.
The following theorem captures the key insight of our algorithm.
\begin{theorem}
		\label{thm:dtmc:prob}
	For MC $\dtmc$ let $G$ and $\StatesMaybe$ be as above and $k\ge0$  with $\Prob_s(\globally^{\le k} \StatesMaybe ) < 1$  for all $s \in \StatesMaybe$. 
	We have
	\begin{align*}
	&\reachprk{k} + \Prob( \globally^{\le k} \StatesMaybe) \cdot \min_{s \in \StatesMaybe} \frac{	\Prob_s(\eventually^{\le k} G) }{1 - \Prob_s(\globally^{\le k} S_{?})} \\
	\le\,  \reachpr
	\le\, &	\reachprk{k} + \Prob( \globally^{\le k} \StatesMaybe) \cdot \max_{s \in \StatesMaybe} \frac{	\Prob_s(\eventually^{\le k} G) }{1 - \Prob_s(\globally^{\le k} S_{?})}.
	\end{align*}
\end{theorem}
	\cref{thm:dtmc:prob} allows us to approximate $\reachpr$ by computing for increasing $k \in \NN$
	\begin{compactitem}
	\item $\reachprk{k}$, the probability to reach a state in $G$ within $k$ steps, and
	\item $\Prob( \globally^{\le k} \StatesMaybe)$, the probability to stay in $\StatesMaybe$ during the first $k$ steps.
	\end{compactitem}
	This can be realized via a value-iteration based procedure.
	The obtained bounds on $\reachpr$ can be tightened arbitrarily since $\Prob(\globally^{\le k} \StatesMaybe)$ approaches 0 for increasing $k$.
	In the following, we address  the correctness of \cref{thm:dtmc:prob}, describe the details of our algorithm, and indicate how the results can be lifted to expected rewards.
	
\subsection{Approximating Reachability Probabilities}
\label{sec:dtmc:correctness}
To approximate the reachability probability $\reachpr$, we consider the step bounded reachability probability $\reachprk{k}$ for $k \ge 0$ and provide a lower and an upper bound for the `missing' probability $\reachpr - \reachprk{k}$.
Note that $ \eventually G$ is the disjoint union of the paths that reach $G$ \emph{within} $k$ steps (given by $\eventually^{\le k} G$) and the paths that reach $G$ only \emph{after} $k$ steps (given by $\StatesMaybe \until^{> k} G$).
\begin{lemma}
	\label{lem:dtmc:probsplit}
For any $k \ge 0$ we have $\reachpr =   \reachprk{k} +  \Prob(\StatesMaybe \until^{> k} G)$. 
\end{lemma}
A path $\pi \in \StatesMaybe \until^{>k} G$ reaches some state $s \in \StatesMaybe$ after \emph{exactly} $k$ steps.
This yields the partition
$\StatesMaybe \until^{>k} G = \bigcupdot_{s \in \StatesMaybe} (\StatesMaybe \until^{=k} \{s\} \cap \eventually G)$.
It follows that
\[
\Prob(\StatesMaybe \until^{> k} G)
= \sum_{s \in \StatesMaybe} \Prob(\StatesMaybe \until^{= k} \{ s\}) \cdot \Prob_s(\eventually G).
\]
Consider $\ell, u \in [0,1]$ with $\ell \le \Prob_s(\eventually G) \le u$ for all $s \in \StatesMaybe$, i.e.,
$\ell$ and $u$ are lower and upper bounds for the reachability probabilities within $\StatesMaybe$. We have
\[
\sum_{s \in \StatesMaybe} \Prob(\StatesMaybe \until^{= k} \{ s\}) \cdot \Prob_s(\eventually G)
\le  \sum_{s \in \StatesMaybe} \Prob(\StatesMaybe \until^{= k} \{ s\}) \cdot u
= \Prob(\globally^{\le k} \StatesMaybe) \cdot u.
\]
We can argue similar for the lower bound $\ell$. With $\cref{lem:dtmc:probsplit}$ we get the following.
\begin{proposition}
	\label{prob:dtmc:reachpr}
	For MC $\dtmc$ with $G$, $\StatesMaybe$, $\ell$, $u$ as above and any $k \ge 0$ we have
	\[
	\reachprk{k} + \Prob(\globally^{\le k} \StatesMaybe) \cdot \ell 
	\le \reachpr
	\le \reachprk{k} + \Prob(\globally^{\le k} \StatesMaybe) \cdot u. 
	\]
\end{proposition}
\begin{remark}
	\label{rem:iisamebounds}
	The bounds for $\reachpr$ given by \cref{prob:dtmc:reachpr} are similar to the bounds obtained after performing $k$ iterations of interval iteration with starting vectors $x_\ell, x_u \in \RR^{|S|}$, where $\vectoraccessSet{x_\ell}{\StatesMaybe} = \ell$ and $\vectoraccessSet{x_u}{\StatesMaybe} = u$.
\end{remark}
We now discuss how the bounds $\ell$ and $u$ can be obtained from the step bounded probabilities $ \Prob_s(\eventually^{\le k} G)$ and $\Prob_s(\globally^{\le k} \StatesMaybe)$ for $s \in \StatesMaybe$.
We focus on the upper bound $u$. 
The reasoning for the lower bound $\ell$ is similar.

Let $\smax \in \StatesMaybe$ be a state with maximal reachability probability, that is $\smax \in \argmax_{s \in \StatesMaybe} \Prob_s(\eventually G)$.
From \cref{prob:dtmc:reachpr} we get
\[
\Prob_\smax(\eventually G) \le \Prob_\smax(\eventually^{\le k} G) + \Prob_\smax(\globally^{\le k} \StatesMaybe) \cdot \Prob_\smax(\eventually G).
\]
We solve the inequality for   $\Prob_\smax(\eventually G)$ (assuming $ \Prob_s(\globally^{\le k}\StatesMaybe) < 1$ for all $s \in \StatesMaybe$):
\[
\Prob_\smax(\eventually G)
\le \frac{\Prob_\smax(\eventually^{\le k} G)}{1-\Prob_\smax(\globally^{\le k} \StatesMaybe)}
\le \max_{s \in \StatesMaybe} \frac{\Prob_s(\eventually^{\le k} G)}{1-\Prob_s(\globally^{\le k} \StatesMaybe)}.
\]
\begin{proposition}
	\label{prob:dtmc:bounds}
	For MC $\dtmc$ let $G$ and $\StatesMaybe$ be as above and $k \ge 0$ such that $ \Prob_s(\globally^{\le k}\StatesMaybe) < 1$ for all $s \in \StatesMaybe$. For every $\hat{s} \in \StatesMaybe$ we have 
 \[
 \min_{s \in \StatesMaybe} \frac{\Prob_s(\eventually^{\le k} G)}{1-\Prob_s(\globally^{\le k} \StatesMaybe)}
 \le \Prob_{\hat{s}}(\eventually G)
 \le 
 \max_{s \in \StatesMaybe} \frac{\Prob_s(\eventually^{\le k} G)}{1-\Prob_s(\globally^{\le k} \StatesMaybe)}.
 \]
\end{proposition}
\cref{thm:dtmc:prob} is a direct consequence of \cref{prob:dtmc:reachpr,prob:dtmc:bounds}.

\subsection{Extending the Value Iteration Approach}
\label{sec:dtmc:algorithm}
Recall the standard value iteration algorithm for approximating $\reachpr$ as discussed in \cref{sec:prelim:modelcheck}.
The function $f \colon \RR^{|S|} \to \RR^{|S|}$ for MCs simplifies to $\vectoraccessSet{f(x)}{\StatesZero} = 0$, $\vectoraccessSet{f(x)}{G} = 1$, and 
$\vectoraccess{f(x)}{s} = \sum_{s' \in S} \probP(s,s') \cdot \vectoraccess{x}{s'}$ for $s \in \StatesMaybe$.
%
We can compute the $k$-step bounded reachability probability  at every state $s \in \States$ by performing $k$ iterations of value iteration~\cite[Remark~10.104]{BK08}.
More precisely, when applying $f$ $k$ times on starting vector $x \in \RR^{|\States|}$ with $\vectoraccessSet{x}{G} = 1$ and $\vectoraccessSet{x}{\States \setminus G} = 0$ we get 
$
 \Prob_s(\eventually^{\le k} G)=\vectoraccess{f^k(x)}{s}.
$
The probabilities $\Prob_s(\globally^{\le k} \StatesMaybe)$ for $s \in \States$ can be computed similarly.
Let $h \colon  \RR^{|S|} \to \RR^{|S|}$ with $\vectoraccessSet{h(y)}{\States \setminus \StatesMaybe} = 0$ and 
$\vectoraccess{h(y)}{s} = \sum_{s' \in S} \probP(s,s') \cdot \vectoraccess{y}{s'}$ for $s \in \StatesMaybe$. 
For starting vector $y \in \RR^{|S|}$ with $\vectoraccessSet{y}{\StatesMaybe} = 1$ and $\vectoraccessSet{y}{\States \setminus \StatesMaybe} = 0$ we get
$\Prob_s(\globally^{\le k }\StatesMaybe) = \vectoraccess{h^k(y)}{s}$.

\cref{alg:dtmc} depicts our approach.
It maintains vectors $x_k,y_k \in \RR^{|\States|}$ which, after $k$ iterations of the loop, store the $k$-step bounded probabilities $\Prob_s(\eventually^{\le k} G)$ and $\Prob_s(\globally^{\le k} \StatesMaybe)$, respectively.
Additionally, the algorithm considers lower bounds $\ell_k$ and upper bounds $u_k$ such that the following invariant holds.
\begin{lemma}
	\label{lem:dtmc:invariant}
	After executing the loop of \cref{alg:dtmc} $k$ times we have for all $s \in \StatesMaybe$ that 
$
	\vectoraccess{x_k}{s} = \Prob_s(\eventually^{\le k} G)$, $\vectoraccess{y_k}{s} = \Prob_s(\globally^{\le k} \StatesMaybe)$, and $\ell_k \le \Prob_s(\eventually G) \le u_k
$.
\end{lemma}
The correctness of the algorithm follows from \cref{thm:dtmc:prob}.
Termination is guaranteed since $\Prob(\eventually (\StatesZero \cup G) ) = 1$ and therefore $\lim_{k\to\infty} \Prob( \globally^{\le k} \StatesMaybe) = \Prob( \globally \StatesMaybe) = 0$.
\begin{theorem}
	\label{thm:dtmc:alg}
	\cref{alg:dtmc} terminates for any MC $\dtmc$, goal states $G$, and precision $\precision > 0$. The returned value $r$ satisfies $|r - \reachpr| < \varepsilon$.
\end{theorem} 

\begin{algorithm}[t]
	\Input{MC $\dtmcDef$, absorbing states $G \subseteq \States$, precision $\precision > 0$}
	\Output{$\result \in \RR$ with $|r - \reachpr| < \precision$}
	$\StatesMaybe \gets \States \setminus \big(\{ s \in \States \mid \Prob_s(\eventually G) = 0  \} \cup G\big)$\\
	initialize $x_0,y_0 \in \RR^{|S|}$ with $\vectoraccessSet{x_0}{G} = 1$, $\vectoraccessSet{x_0}{\States \setminus G} = 0$, $\vectoraccessSet{y_0}{\StatesMaybe} = 1$, $\vectoraccessSet{y_0}{\States \setminus \StatesMaybe} = 0$\\
	$\ell_0 \gets -\infty$; $u_0 \gets +\infty$\\
	$k \gets 0$\\
	\Repeat{$\vectoraccess{y_k}{\sinit} \cdot (u_k-\ell_k) < 2 \cdot \precision$}{%
		$k \gets k+1$\\
		$x_k \gets f(x_{k-1})$; $y_k \gets h(y_{k-1})$\\
		\If{$\vectoraccess{y_k}{s} < 1$ for all $s \in \StatesMaybe$}{%
			$\ell_k \gets \max(\ell_{k-1}, \min_{s \in \StatesMaybe} \frac{\vectoraccess{x_k}{s}}{1-\vectoraccess{y_k}{s}})$;
			$u_k \gets \min(u_{k-1}, \max_{s \in \StatesMaybe} \frac{\vectoraccess{x_k}{s}}{1-\vectoraccess{y_k}{s}})$
			}
	}
	\Return{$\vectoraccess{x_k}{\sinit} + \vectoraccess{y_k}{\sinit} \cdot \frac{\ell_k + u_k}{2}$ }
	\caption{Sound value iteration for MCs.}
	\label{alg:dtmc}
\end{algorithm}

\begin{example}
	We apply \cref{alg:dtmc} for the MC in \cref{fig:prelim:dtmc} and the set of goal states $G = \{s_4\}$.
	We have $\StatesMaybe = \{s_0, s_1, s_2\}$.
	After $k=3$ iterations it holds that
	\begin{align*}
		\vectoraccess{x_3}{s_0} = 0.00003 \quad 
		\vectoraccess{x_3}{s_1} &= 0.003 \quad 
		\vectoraccess{x_3}{s_2} = 0.3 \\ 
		\vectoraccess{y_3}{s_0} = 0.99996 \quad 
		\vectoraccess{y_3}{s_1} &= 0.996 \quad 
		\vectoraccess{y_3}{s_2} = 0.6 
	\end{align*}
	Hence, $\frac{\vectoraccess{x_3}{s}}{1 - \vectoraccess{y_3}{s}} = \frac{3}{4} = 0.75$ for all $s \in \StatesMaybe$.
	We get $\ell_3 = u_3 = 0.75$.
	The algorithm converges for any $\precision > 0$ and returns the correct solution $\vectoraccess{x_3}{s_0} + \vectoraccess{y_3}{s_0} \cdot 0.75 = 0.75$.
\end{example}

\subsection{Sound Value Iteration for Expected Rewards}
We lift our approach to expected rewards in a straightforward manner.
Let $G \subseteq \States$ be a set of absorbing goal states of MC $\dtmc$ such that\linebreak$\reachpr = 1$. Further let $\StatesMaybe = \States \setminus G$.
For $k\ge 0$ we observe that the expected reward $\exprew$ can be split into the expected reward collected within $k$ steps and the expected reward collected only after $k$ steps, i.e.,
$
\exprew = \exprewk{k} + \sum_{s \in \StatesMaybe} \Prob(\StatesMaybe \until^{=k} \{s\}) \cdot \Exp_s (\eventuallyrew G).
$
Following a similar reasoning as in \cref{sec:dtmc:correctness} we can show the following.
\begin{theorem}
	\label{thm:dtmc:rew}
	For MC $\dtmc$ let $G$ and $\StatesMaybe$ be as before and $k\ge0$  such that $\Prob_s(\globally^{\le k} \StatesMaybe ) < 1$  for all $s \in \StatesMaybe$. 
	We have
	\begin{align*}
	&\exprewk{k} + \Prob( \globally^{\le k} \StatesMaybe) \cdot \min_{s \in \StatesMaybe} \frac{	\Exp_s(\eventuallyrew^{\le k} G) }{1 - \Prob_s(\globally^{\le k} S_{?})} \\
	\le\,  \exprew
	\le\, &	\exprewk{k} + \Prob( \globally^{\le k} \StatesMaybe) \cdot \max_{s \in \StatesMaybe} \frac{	\Exp_s(\eventuallyrew^{\le k} G) }{1 - \Prob_s(\globally^{\le k} S_{?})}.
	\end{align*}
\end{theorem}
Recall the function $g \colon \RR^{|\States|} \to \RR^{|\States|}$ from \cref{sec:prelim:modelcheck}, given by $\vectoraccessSet{g(x)}{G} = 0$ and 
$
\vectoraccess{g(x)}{s} = \rewFct(s) + \sum_{s' \in \States} \probP(s, s') \cdot \vectoraccess{x}{s'}
$
for $s \in \StatesMaybe$.
For $s \in \States$ and $x \in \RR^{|\States|}$ with $\vectoraccessSet{x}{\States} = 0$ we have $\Exp_s(\eventuallyrew^{\le k} G) = \vectoraccess{g^k(x)}{s}$.
We modify \cref{alg:dtmc} such that it considers function $g$ instead of function $f$.
Then, the returned value $\result$ satisfies $|r - \exprew| < \precision$.

\subsection{Optimizations.}\label{sec:dtmc:opt}
\Cref{alg:dtmc} can make use of \emph{initial bounds} $\ell_0, u_0 \in \RR$ with $\ell_0 \le \Prob_s(\eventually G) \le u_0$ for all $s \in \StatesMaybe$.
Such bounds could be derived, e.g., from domain knowledge or during preprocessing~\cite{DBLP:conf/cav/Baier0L0W17}.
The algorithm always chooses the largest available lower bound for $\ell_k$  and the lowest available upper bound for $u_k$, respectively.
If \cref{alg:dtmc} and interval iteration are initialized with the same bounds, \cref{alg:dtmc} always requires as most as many iterations compared to interval iteration (cf. \cref{rem:iisamebounds}).

\emph{Gauss-Seidel value iteration}~\cite{Put94,DBLP:conf/cav/Baier0L0W17} is an optimization for standard value iteration and interval iteration that potentially leads to faster convergence.
When computing $\vectoraccess{f(x)}{s}$ for $s \in \StatesMaybe$, the idea is to consider already computed results $\vectoraccess{f(x)}{s'}$ from the current iteration.
Formally, let ${\prec} \subseteq \States \times \States$ be some strict total ordering of the states.
Gauss-Seidel value iteration considers instead of function $f$ the function $f_\prec \colon \RR^{|\States|} \to \RR^{|\States|}$ with 
$\vectoraccessSet{f_\prec}{\StatesZero} = 0$, $\vectoraccessSet{f_\prec}{G} = 1$, and 
\[
\vectoraccess{f_\prec(x)}{s} =
\sum_{s' \prec s} \probP(s, s') \cdot \vectoraccess{f_\prec(x)}{s'} +
\sum_{s' \not\prec s} \probP(s, s') \cdot \vectoraccess{x}{s'}.
\]
Values $\vectoraccess{f_\prec(x)}{s}$ for $s \in \States$ are computed in the order defined by $\prec$.
This idea can also be applied to our approach.
To this end, we replace $f$ by $f_\prec$ and $h$ by $h_\prec$, where $h_\prec$ is defined similarly.
More details are given in \cref{app:gs}.

\emph{Topological value iteration}~\cite{DBLP:journals/jair/DaiMWG11} employs the graphical structure of the MC $\dtmc$.
The idea is to decompose the states $\States$ of $\dtmc$ into strongly connected components\footnote{$\States' \subseteq \States$ is a connected component if $s$ can be reached from $s'$ for all $s,s' \in \States'$. $\States'$ is a strongly connected component if no superset of $\States'$ is a connected component.} (SCCs) that are analyzed individually.
The procedure can improve the runtime of classical value iteration since for a single iteration only the values for the current SCC have to be updated.
A topological variant of interval iteration is introduced in \cite{DBLP:conf/cav/Baier0L0W17}.
Given these results, sound value iteration can be extended similarly.
  \section{Sound Value Iteration for MDPs}
We extend sound value iteration to compute reachability probabilities in MDPs.
Assume an MDP $\mdpDef$ and a set of absorbing goal states $G$.
For simplicity, we focus on maximal reachability probabilities, i.e., we compute $\reachprSched[\mathrm{max}]$. Minimal reachability probabilities and expected rewards are analogous.
As in \cref{sec:prelim:modelcheck} we consider the partition $\States = \StatesZero \cupdot G \cupdot \StatesMaybe$ such that $\mdp$ is contracting with respect to $G \cup \StatesZero$.

\subsection{Approximating Maximal Reachability Probabilities}
We argue that our results for MCs also hold for MDPs under a given scheduler $\sched \in \Scheds[\mdp]$.
Let $k\ge0$  such that $\ProbSched_s(\globally^{\le k} \StatesMaybe ) < 1$  for all $s \in \StatesMaybe$.
Following the reasoning as in \cref{sec:dtmc:correctness} we get
	\begin{align*}
	\reachprkSched{k} + \ProbSched( \globally^{\le k} \StatesMaybe) \cdot \min_{s \in \StatesMaybe} \frac{	\ProbSched_s(\eventually^{\le k} G) }{1 - \ProbSched_s(\globally^{\le k} \StatesMaybe)} 	\le  \reachprSched 
	\le \reachprSched[\mathrm{max}].
	\end{align*}
Next, assume an upper bound $u \in \RR$ with $\ProbSched[\mathrm{max}]_s(\eventually G) \le u$ for all $s \in \StatesMaybe$.
For a scheduler $\schedmax \in \Scheds[\mdp]$ that attains the maximal reachability probability, i.e.,
$\schedmax \in \arg\max_{\sched \in \Scheds[\mdp]} \ProbSched(\eventually G)$ it holds that
\begin{align*}
\reachprSched[\mathrm{max}]
= \ProbSched[\schedmax](\eventually G)
& \le	\reachprkSched[\schedmax]{k} + \ProbSched[\schedmax]( \globally^{\le k} \StatesMaybe) \cdot u\\
& \le	\max_{\sched \in \Scheds[\mdp]} \big( \reachprkSched{k} + \ProbSched( \globally^{\le k} \StatesMaybe) \cdot u\big).
\end{align*}
We obtain the following theorem which is the basis of our algorithm.
\begin{theorem}
	\label{thm:mdp:bounds}
	For MDP $\mdp$ let $G$, $\StatesMaybe$, and $u$ be as above. 
	Assume  $\sched \in \Scheds[\mdp]$ and  $k \ge 0$ such that $\sched \in \argmax_{\sched' \in \Scheds[\mdp]} \ProbSched[\sched'](\eventually^{\le k} G) + \ProbSched[\sched']( \globally^{\le k} \StatesMaybe) \cdot u$  and $\ProbSched_s(\globally^{\le k} \StatesMaybe ) < 1$ for all $s \in \StatesMaybe$.
	We have 
 	\begin{align*}
 	&\reachprkSched{k} + \ProbSched( \globally^{\le k} \StatesMaybe) \cdot \min_{s \in \StatesMaybe} \frac{	\ProbSched_s(\eventually^{\le k} G) }{1 - \ProbSched_s(\globally^{\le k} \StatesMaybe)}\\
 	\le\,&  \reachprSched[\mathrm{max}]
 	\le \reachprkSched{k} + \ProbSched( \globally^{\le k} \StatesMaybe) \cdot u.
 \end{align*}
\end{theorem}
Similar to the results for MCs it also holds that 
$\reachprSched[\mathrm{max}] \le  \max_{\sched \in \Scheds[\mdp]} \hat{u}_k^\sched$ with
\[
\hat{u}_k^\sched \coloneqq \reachprkSched{k} + \ProbSched( \globally^{\le k} \StatesMaybe) \cdot \max_{s \in \StatesMaybe} \frac{	\ProbSched_s(\eventually^{\le k} G) }{1 - \ProbSched_s(\globally^{\le k} \StatesMaybe)}.
\]
However, this upper bound can not trivially be embedded in a value iteration based procedure.
Intuitively, in order to compute the upper bound for iteration $k$, one can not necessarily build on the results for iteration $k-1$.

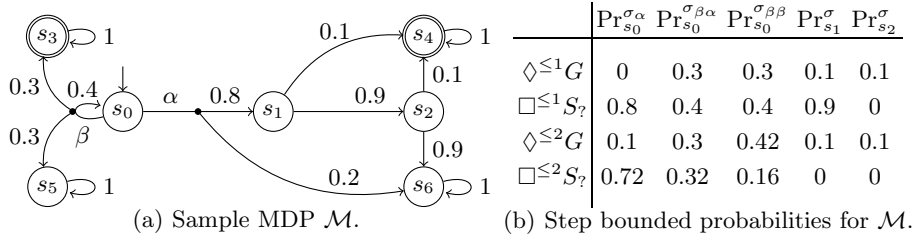
\begin{figure}[t]
	\centering
	\subfigure[Sample MDP $\mdp$.]{
		\label{fig:mdp:ex}
		\begin{tikzpicture}[scale=1,baseline=(current bounding box.east)]
		\node [state] (s0) at (0,0) {$s_0$};
		\node [state] (s1) [on grid, right =20mm of s0] {$s_1$};
		\node [state] (s2) [on grid, right =20mm of s1] {$s_2$};
		\node [state, target] (s4) [on grid, above=10mm of s2] {$s_4$};
		\node [state, target] (s3) [on grid, left=50mm of s4] {$s_3$};
		\node [state] (s6) [on grid, below=10mm of s2] {$s_6$};
		\node [state] (s5) [on grid, left=50mm of s6] {$s_5$};
		
		\initstateAbove{s0}

		\path
			(s0) edge[transition] node[pos=0.25,above] {$\act$} node[action] (alpha) {} node[pos=0.75,above] {$0.8$} (s1)
			(s0) edge[transition, loop left] node[pos=0.25,below] {$\altact$} node[action] (beta) {} node[pos=0.75,above] {$0.4$} (s0)
			(beta) edge[transition, bend left] node[left] {$0.3$} (s3)
			(beta) edge[transition, bend right] node[left] {$0.3$} (s5)
			(alpha) edge[transition, bend right] node[above,pos=0.75] {$0.2$} (s6)
			(s1) edge[transition] node[above, pos=0.7] {$0.9$} (s2)
			(s1) edge[transition, bend left] node[above] {$0.1$} (s4)
			(s2) edge[transition] node[right] {$0.1$} (s4)
			(s2) edge[transition] node[right] {$0.9$} (s6)
			(s3) edge[transition, loop right] node [right] {$1$} (s3)
			(s4) edge[transition, loop right] node [right] {$1$} (s4)
			(s5) edge[transition, loop right] node [right] {$1$} (s5)
			(s6) edge[transition, loop right] node [right] {$1$} (s6)
		; 
		\end{tikzpicture}}%
	\hfill
	\subfigure[Step bounded probabilities for $\mdp$.]{
		\label{fig:mdp:table}
		\renewcommand{\arraystretch}{1.2}%
	\begin{tabular}{c|ccccc}
		& $\ProbSched[\sched_{\act}]_{s_0}$ & $\ProbSched[\sched_{\altact\act}]_{s_0}$ & $\ProbSched[\sched_{\altact\altact}]_{s_0}$ & $\ProbSched_{s_1}$ & $\ProbSched_{s_2}$ \\ \hline\\[-7pt]
		$\eventually^{\le 1} G$          & 0                                 & 0.3                                      & 0.3                                         & 0.1                & 0.1                \\
		$\globally^{\le 1} \StatesMaybe$ & 0.8                               & 0.4                                      & 0.4                                         & 0.9                & 0                  \\
		$\eventually^{\le 2} G$          & 0.1                               & 0.3                                      & 0.42                                        & 0.1                & 0.1                \\
		$\globally^{\le 2} \StatesMaybe$ & 0.72                              & 0.32                                     & 0.16                                        & 0                  & 0                 \\[5pt]
	\end{tabular}
	}
	\caption{Example MDP with corresponding step bounded probabilities.}
\end{figure}

\begin{example}
	\label{ex:mdp:nontrivialupperbound}
	Consider the MDP $\mdp$ given in \cref{fig:mdp:ex}. Let $G = \{s_3,s_4\}$ be the set of goal states. We therefore have $\StatesMaybe = \{s_0, s_1, s_2\}$.
	In \cref{fig:mdp:table} we list step bounded probabilities with respect to the possible schedulers, where  $\sched_\act$, $\sched_{\altact\act}$, and $\sched_{\altact\altact}$ refer to schedulers with
	$\sched_\act(s_0) = \act$ and for $\gamma \in \{\act,\altact\}$, $\sched_{\altact\gamma}(s_0) =   \altact$ and $\sched_{\altact\gamma}(s_0\altact s_0) = \gamma$.
	Notice that the probability measures $\ProbSched_{s_1}$ and $\ProbSched_{s_2}$ are independent of the considered scheduler $\sched$.
	For step bounds $k \in \{1,2\}$ we get
	\begin{compactitem}
	\item $\max_{\sched \in \Scheds[\mdp]} \hat{u}_1^\sched =
	\hat{u}_1^{\sched_\act} = 0 + 0.8 \cdot \max(0,1,0)  = 0.8$ and
	\item $\max_{\sched \in \Scheds[\mdp]} \hat{u}_2^\sched =
	\hat{u}_2^{\sched_{\altact\altact}} = 0.42+0.16 \cdot  \max(0.5,0.19,1) = 0.5.$
	\end{compactitem}
\end{example}
\begin{algorithm}[t]
	\Input{MDP $\mdpDef$, absorbing states $G \subseteq \States$, precision $\precision > 0$}
	\Output{$\result \in \RR$ with $|r - \reachprSched[\mathrm{max}]| < \precision$}
	$\StatesZero \gets \{ s \in \States \mid \ProbSched[\mathrm{max}]_s(\eventually G) = 0  \}$\\
	assert that $\mdp$ is contracting with respect to $G \cup \StatesZero$\\
	$\StatesMaybe \gets \States \setminus (\StatesZero \cup G)$\\
	initialize $x_0,y_0 \in \RR^{|S|}$ with $\vectoraccessSet{x_0}{G} = 1$, $\vectoraccessSet{x_0}{\States \setminus G} = 0$, $\vectoraccessSet{y_0}{\StatesMaybe} = 1$, $\vectoraccessSet{y_0}{\States \setminus \StatesMaybe} = 0$\\
	$\ell_0 \gets -\infty$; $u_0\gets +\infty$; $d_0 \gets -\infty$\\
	$k \gets 0$\\
	\Repeat{$\vectoraccess{y_k}{\sinit} \cdot (u_k-\ell_k) < 2 \cdot \precision$}{%
		$k \gets k + 1$\\
		initialize $x_k,y_k \in \RR^{|S|}$ with $\vectoraccessSet{x_k}{G} = 1$, $\vectoraccessSet{x_k}{\StatesZero} = 0$, $\vectoraccessSet{y_k}{\States \setminus \StatesMaybe} = 0$\\
		$d_k \gets d_{k-1}$\\
		\ForEach{$s \in \StatesMaybe$}{%
			$\act \gets \mathit{findAction}(x_{k-1},y_{k-1},s,u_{k-1})$\\
			$d_k \gets \max(d_k, \mathit{decisionValue}(x_{k-1},y_{k-1},s,\act))$\\
			$\vectoraccess{x_k}{s} \gets  \sum_{s' \in S} \probP(s, \act, s') \cdot \vectoraccess{x_{k-1}}{s'}$\\ 
			$\vectoraccess{y_k}{s} \gets  \sum_{s' \in S} \probP(s, \act, s') \cdot \vectoraccess{y_{k-1}}{s'}$\label{line:mdp:vectors}\\ 
		}
		\If{$\vectoraccess{y_k}{s} < 1$ for all $s \in \StatesMaybe$}{%
			$\ell_k \gets \max(\ell_{k-1}, \min_{s \in \StatesMaybe} \frac{\vectoraccess{x_k}{s}}{1-\vectoraccess{y_k}{s}})$\\
			$u_k \gets \min(u_{k-1}, \max(d_{k}, \max_{\state \in \StatesMaybe} \frac{\vectoraccess{x_k}{s}}{1-\vectoraccess{y_k}{s}} ))$\\
			\label{line:mdp:updateu}
		}
	}
	\Return{$\vectoraccess{x_k}{\sinit} + \vectoraccess{y_k}{\sinit} \cdot \frac{\ell_k + u_k}{2}$ }\\[12pt]
	\caption{Sound value iteration for MDPs}
	\label{alg:mdp}
\end{algorithm}
\subsection{Extending the Value Iteration Approach}
The idea of our algorithm is to compute  the bounds for $\reachprSched[\mathrm{max}]$ as in \cref{thm:mdp:bounds} for increasing $k \ge 0$.
\cref{alg:mdp} outlines the procedure.
Similar to \cref{alg:dtmc} for MCs, vectors $x_k,y_k \in \RR^{|\States|}$ store the step bounded probabilities $\ProbSched[\sched_k]_s(\eventually^{\le k} G)$ and $\ProbSched[\sched_k]_s(\globally^{\le k} \StatesMaybe)$ for any $s \in \States$.
In addition, schedulers $\sched_k$ and upper bounds $u_k \ge \max_{\state \in \StatesMaybe} \ProbSched[\mathrm{max}]_s(\eventually G)$ are computed in a way that \cref{thm:mdp:bounds} is applicable.
%
\begin{lemma}
	\label{lem:mdpalginvariant}
	After executing $k$ iterations of \cref{alg:mdp} we have for all $s \in \StatesMaybe$ that
$
	\vectoraccess{x_k}{s} = \ProbSched[\sched_k]_s(\eventually^{\le k} G)$,
	$\vectoraccess{y_k}{s} = \ProbSched[\sched_k]_s(\globally^{\le k} \StatesMaybe)$, and 
	$\ell_k  \le \ProbSched[\mathrm{max}]_s(\eventually G) \le u_k$,
   where $\sched_k \in \argmax_{\sched \in \Scheds[\mdp]} \ProbSched_s(\eventually^{\le k} G) + \ProbSched_s( \globally^{\le k} \StatesMaybe) \cdot u_k$.
\end{lemma}
%
%
%
The lemma holds for $k=0$ as $x_0$, $y_0$, and $u_0$ are initialized accordingly.
For $k>0$ we assume that the claim holds after $k-1$ iterations, i.e., for $x_{k-1}$, $y_{k-1}$, $u_{k-1}$ and scheduler $\sched_{k-1}$.
The results of the $k$th iteration are obtained as follows.
%

\begin{algorithm}[t]
	\Function{$\mathit{findAction}(x,y,s,u)$}{%
		\If{$u \neq \infty$}{%
			\Return $\act \in  \argmax_{\act \in \Act{s}}  \sum_{s' \in S} \probP(s, \act, s') \cdot (\vectoraccess{x}{s'} + \vectoraccess{y}{s'} \cdot u)$
		}\Else{
			\Return $\act \in \argmax_{\act \in \Act{s}}  \sum_{s' \in S} \probP(s, \act, s') \cdot (\vectoraccess{y}{s'})$
		}
	}
	\caption{Computation of optimal action.}
	\label{alg:mdp:findAction}
\end{algorithm}

The function $\mathit{findAction}$ illustrated in \cref{alg:mdp:findAction} determines the choices of a scheduler $\sched_k \in \argmax_{\sched \in \Scheds[\mdp]} \ProbSched_s(\eventually^{\le k} G) + \ProbSched_s( \globally^{\le k} \StatesMaybe) \cdot u_{k-1}$ for $s \in \StatesMaybe$.
The idea is to consider at state $s$ an action $\sched_k(s) = \act \in \Act{s}$ that maximizes 
\begin{align*}
  \ProbSched[\sched_k]_s(\eventually^{\le k} G) + \ProbSched[\sched_k]_s(\globally^{\le k} \StatesMaybe) \cdot u_{k-1} 
 =\! \sum_{s' \in \States}\! \probP(s, \act, s')\! \cdot\! (\vectoraccess{x_{k-1}}{s'} + \vectoraccess{y_{k-1}}{s'} \cdot u_{k-1}).
\end{align*}
For the case where no real upper bound is known (i.e., $u_{k-1} = \infty$) we implicitly assume a sufficiently large value for $u_{k-1}$ such that $\ProbSched_s(\eventually^{\le k} G)$ becomes negligible.
Upon leaving state $s$, $\sched_k$ mimics $\sched_{k-1}$, i.e., we set $\sched_k(s \act s_1 \act_1 \dots s_n) = \sched_{k-1}(s_1 \act_1 \dots s_n)$.
After executing \cref{line:mdp:vectors} of \cref{alg:mdp} we have 
$\vectoraccess{x_k}{s} = \ProbSched[\sched_k]_s(\eventually^{\le k} G)$ and $\vectoraccess{y_k}{s} = \ProbSched[\sched_k]_s(\globally^{\le k} \StatesMaybe)$.
It remains to derive an upper bound $u_{k}$. 
To ensure that \cref{lem:mdpalginvariant} holds we require (i) $u_k \ge \max_{s \in \StatesMaybe} \ProbSched[\mathrm{max}]_s (\eventually G)$ and (ii) $u_k \in U_k$, where 
\[
U_k = \{ u \in \RR \mid \sched_k \in \argmax_{\sched \in \Scheds[\mdp]} \ProbSched_s(\eventually^{\le k} G) + \ProbSched_s(\globally^{\le k} \StatesMaybe) \cdot u \text{ for all } s \in \StatesMaybe \}.
\]
Intuitively, the set $U_k \subseteq \RR$ consists of all possible upper bounds $u$ for which $\sched_k$ is still optimal.
$U_k \subseteq$ is convex as it can be represented as a conjunction of inequalities with $U_0 = \RR$ and $u \in U_k$ if and only if $u \in U_{k-1}$ and for all $s \in \StatesMaybe$ with $\sched_k(s) = \act$ and for all $\altact \in \Act{s} \setminus \{\act\}$
\begin{align*}
\sum_{s' \in \States}\! \probP(s, \act, s')\! \cdot\! (\vectoraccess{x_{k-1}}{s'} + \vectoraccess{y_{k-1}}{s'} \cdot u)
\ge \!
\sum_{s' \in \States}\! \probP(s, \altact, s')\! \cdot\! (\vectoraccess{x_{k-1}}{s'} + \vectoraccess{y_{k-1}}{s'} \cdot u).
\end{align*}
The algorithm maintains the so-called \emph{decision value} $d_k$ which corresponds to the minimum of $U_k$ (or $-\infty$ if the minimum does not exist).
\cref{alg:mdp:decisionvalue} outlines the procedure to obtain the decision value at a given state.
Our algorithm ensures that $u_k$ is only set to a value in $[d_k, u_{k-1}] \subseteq U_k$.

\begin{lemma}
After executing \cref{line:mdp:updateu} of \cref{alg:mdp}:  $u_k \ge \max_{s \in \StatesMaybe} \ProbSched[\mathrm{max}]_s (\eventually G)$.
	
\end{lemma}
To show that $u_k$ is a valid upper bound, let $\smax \in \argmax_{s \in \StatesMaybe}\ProbSched[\mathrm{max}]_s(\eventually G)$  and $u^* = \ProbSched[\mathrm{max}]_\smax(\eventually G)$.
From \cref{thm:mdp:bounds}, $u_{k-1} \ge u^*$, and $u_{k-1} \in U_k$ we get
\begin{align*}
u^*
& \le
\max_{\sched \in \Scheds[\mdp]}\ProbSched[\sched]_\smax(\eventually^{\le k} G) + \ProbSched[\sched]_\smax(\globally^{\le k} \StatesMaybe) \cdot u_{k-1}\\
& = \ProbSched[\sched_k]_\smax(\eventually^{\le k} G) + \ProbSched[\sched_k]_\smax(\globally^{\le k} \StatesMaybe) \cdot u_{k-1}
 = \vectoraccess{x_{k}}{\smax} + \vectoraccess{y_k}{\smax} \cdot u_{k-1}
\end{align*}
which yields a new upper bound $\vectoraccess{x_{k}}{\smax} + \vectoraccess{y_k}{\smax} \cdot u_{k-1} \ge u^*$.
We repeat this scheme as follows. Let $v_0 \coloneqq u_{k-1}$ and for $i > 0$ let $v_i \coloneqq \vectoraccess{x_{k}}{\smax} + \vectoraccess{y_k}{\smax} \cdot v_{i-1}$.
We can show that $v_{i-1} \in U_k$ implies $v_{i} \ge u^*$.
Assuming $\vectoraccess{y_k}{\smax} < 1$, the sequence $v_0, v_1, v_2, \dots$ converges to 
$
v_\infty \coloneqq \lim_{i \to \infty} v_i  = \frac{\vectoraccess{x_k}{\smax}}{1-\vectoraccess{y_k}{\smax}}.
$
We distinguish three cases to show that $u_k = \min(u_{k-1}, \max(d_{k}, \max_{\state \in \StatesMaybe} \frac{\vectoraccess{x_k}{s}}{1-\vectoraccess{y_k}{s}} )) \ge u^*$.
\begin{compactitem}
	\item If $v_\infty > u_{k-1}$, then also $\max_{\state \in \StatesMaybe} \frac{\vectoraccess{x_k}{s}}{1-\vectoraccess{y_k}{s}} > u_{k-1}$. Hence $u_k = u_{k-1} \ge u^*$.
	\item If $d_k \le v_\infty \le u_{k-1}$, 
	we can show that $v_i \le v_{i-1}$.
	It follows that for all $i > 0$, $v_{i-1} \in U_k$, implying $v_{i} \ge u^*$.
	Thus we get $u_k = \max_{\state \in \StatesMaybe} \frac{\vectoraccess{x_k}{s}}{1-\vectoraccess{y_k}{s}}
	\ge v_\infty \ge  u^*$.
	\item If $v_\infty < d_k$ then there is an $i \ge 0$ with $v_i \ge d_k$ and $u^* \le v_{i+1} < d_k$. It follows that $u_k = d_k \ge u^*$.
\end{compactitem}
\begin{algorithm}[t]
	\Function{$\mathit{decisionValue}(x,y,s,\act)$}{%
		$d \gets -\infty$\\
		\ForEach{$\beta \in \Act{s} \setminus \{\act\}$}{%
			$y_\Delta \gets \sum_{s' \in S} (\probP(s, \act, s') - \probP(s, \beta, s')) \cdot \vectoraccess{y}{s'}$\\
			
			\If{$y_\Delta > 0$}{%
				$x_\Delta \gets \sum_{s' \in S} (\probP(s, \beta, s') - \probP(s, \act, s')) \cdot \vectoraccess{x}{s'}$\\
				$d \gets \max(d,  \nicefrac{x_\Delta}{y_\Delta}  )$	
			}
		}
		\Return $d$
		\caption{Computation of decision value.}
		\label{alg:mdp:decisionvalue}
	}
\end{algorithm}

\begin{example}
	Reconsider the MDP $\mdp$ from \cref{fig:mdp:ex} and goal states $G = \{s_3, s_4\}$.
	The maximal reachability probability is attained for a scheduler that always chooses $\altact$ at state $s_0$, which results in $\reachprSched[\mathrm{max}] = 0.5$.
	We now illustrate how \cref{alg:mdp} approximates this value by sketching the first two iterations.
	For the first iteration $\mathit{findAction}$ yields action $\act$ at $s_0$. We obtain:%
	\begin{align*}
	&\vectoraccess{x_1}{s_0} = 0, \ 
	\vectoraccess{x_1}{s_1} = 0.1, \ 
	\vectoraccess{x_1}{s_2} = 0.1, \ 
	\vectoraccess{y_1}{s_0} = 0.8, \ 
	\vectoraccess{y_1}{s_1} = 0.9, \  
	\vectoraccess{y_1}{s_2} = 0,\\
	&d_1 = 0.3/(0.8-0.4) = 0.75, \ 
	\ell_1 = \min(0,1,0) = 0, \
	u_1 = \max(0.75,0,1,0) = 1.
	\end{align*}%
	In the second iteration $\mathit{findAction}$ yields again $\act$ for $s_0$ and we get:
	\begin{align*}
	&\vectoraccess{x_2}{s_0} = 0.08, \ 
	\vectoraccess{x_2}{s_1} = 0.19, \ 
	\vectoraccess{x_2}{s_2} = 0.1, \ 
	\vectoraccess{y_2}{s_0} = 0.72, \ 
	\vectoraccess{y_2}{s_1} = 0, \
	\vectoraccess{y_2}{s_2} = 0,\\
	&d_2 = 0.75, \, 
	\ell_2 = \min(0.29,0.19,0.1) = 0.1, \, 
	u_2 = \max(0.75,0.29,0.19,0.1) = 0.75.
	\end{align*}
	Due to the decision value we do not set the upper bound $u_2$ to $0.29 < \reachprSched[\mathrm{max}]$.
\end{example}

\begin{theorem}
	\label{thm:mdp:alg}
	\cref{alg:mdp} terminates for any MDP $\mdp$, goal states $G$ and precision $\varepsilon > 0$. The returned value $r$ satisfies $|r -\reachprSched[\mathrm{max}]| \le \varepsilon$.
\end{theorem}
The correctness of the algorithm follows from \cref{thm:mdp:bounds} and \cref{lem:mdpalginvariant}.
Termination follows since $\mdp$ is contracting with respect to $\StatesZero \cup G$, implying\linebreak$\lim_{k\to\infty} \ProbSched[\sched](\globally^{\le k} \StatesMaybe) = 0$.
The optimizations for \cref{alg:dtmc} mentioned in \cref{sec:dtmc:opt} can be applied to \cref{alg:mdp} as well.
  \section{Experimental Evaluation}
\label{sec:experiments}
\begin{figure}[t]
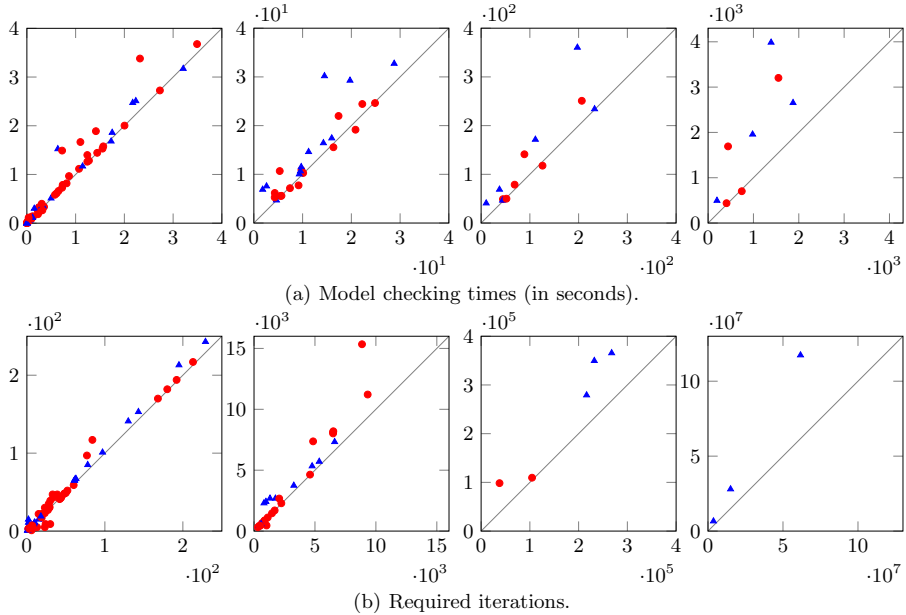

	\centering
	\adjustbox{max width=\linewidth}{
	\subfigure[Model checking times (in seconds).]{%
		\label{fig:experiments:time}
		\setplotlimits{0}{4}{-1}{40}{-2}{400}{-3}{4300}\scatterplot{SVI-time}{II-time}
}}\\[-3pt]
	\adjustbox{max width=\linewidth}{
	\subfigure[Required iterations.]{%
		\label{fig:experiments:iters}
		\setplotlimits{-2}{250}{-3}{16000}{-5}{400000}{-7}{130000000}\scatterplot{SVI-iters}{II-iters}
}}
\iftoggle{tr}{}{\vspace{-15pt}}
	\caption{Comparison of sound value iteration (x-axis) and interval iteration (y-axis).}
	\label{fig:experiments:runtimeiters}
\end{figure}

\subsubsection{Implementation.}
We implemented sound value iteration for MCs and MDPs into the model checker \storm \cite{DBLP:conf/cav/DehnertJK017}.
The implementation computes reachability probabilities and expected rewards using explicit data structures such as sparse matrices and vectors.
Moreover, Multi-objective model checking is supported, where we straightforwardly extend the value iteration-based approach of \cite{ForejtKPatva12} to sound value iteration.
We also implemented the optimizations given in \cref{sec:dtmc:opt}.

The implementation is available at \url{www.stormchecker.org}.

\subsubsection{Experimental Results.}
We considered a wide range of case studies including
\begin{compactitem}
	\item all MCs, MDPs, and CTMCs from the \prism benchmark suite~\cite{KNP12b},
	\item several case studies from the \prism website \url{www.prismmodelchecker.org},
	\item Markov automata accompanying \imca~\cite{DBLP:conf/atva/GuckTHRS14}, and
	\item multi-objective MDPs considered in~\cite{ForejtKPatva12}.
\end{compactitem}
In total, 130 model and property instances were considered.
For CTMCs and Markov automata we computed (untimed) reachability probabilities or expected rewards  on the underlying MC and the underlying MDP, respectively.
In all experiments the precision parameter was given by $\precision = 10^{-6}$.

We compare sound value iteration ($\SVI$) with interval iteration ($\II$) as presented in~\cite{HADDAD2017,DBLP:conf/cav/Baier0L0W17}.
We consider the Gauss-Seidel variant of the approaches and  compute initial bounds $\ell_0$ and $u_0$ as in \cite{DBLP:conf/cav/Baier0L0W17}.
For a better comparison we consider the implementation of $\II$ in \storm.
\Cref{app:experiments} gives a comparison with the implementation of $\II$ in \prism.
The experiments were run on a single core (2GHz) of an HP BL685C G7 with 192GB of available memory. 
However, almost all experiments required less than 4GB.
We measured model checking times and required iterations.
All logfiles and considered benchmarks are available at~\cite{modelslogs}.
\begin{figure}[t]	\pgfplotsset{compat = 1.3}
	\begin{tikzpicture}
	\begin{axis}[	
	ymin=0.3,
	ylabel shift=0pt,
	ymode = log,
	ymax=3900,
	xmin = 50,
	xmax=130,
	width=\textwidth,
	height=7.5cm,
	xlabel=Number of solved instances,
	ylabel=Time (seconds),
	legend style={at={(0.5,0.95)}, anchor=north, legend columns=-1,/tikz/every even column/.append style={column sep=0.3cm}},
	reverse legend,
	]
	
	\addplot[only marks, mark size=1.3pt, gray!70, mark=*] table [x=num,y=VI, col sep=semicolon] {plotdata/alltimes.csv};
	\addplot[only marks, violet, mark size = 1.4pt, mark=triangle] table [x=num,y=topo-QVI, col sep=semicolon] {plotdata/alltimes.csv};
	\addplot[only marks, Green, mark size=1.3pt, mark=o] table [x=num,y=topo-sym-II, col sep=semicolon] {plotdata/alltimes.csv};
	\addplot[only marks, blue, mark=x] table [x=num,y=QVI, col sep=semicolon] {plotdata/alltimes.csv};
	\addplot[only marks, red, mark=+] table [x=num,y=sym-II, col sep=semicolon] {plotdata/alltimes.csv};
	\legend{$\VI$, topol.\,$\SVI$ , topol.\,$\II$, $\SVI$ ,$\II$ }
	\end{axis}
	\end{tikzpicture}
	
\iftoggle{tr}{}{\vspace{-10pt}}
	\caption{Runtime comparison between different approaches.}
\iftoggle{tr}{}{\vspace{-10pt}}
	\label{fig:experiments:all}
\end{figure}
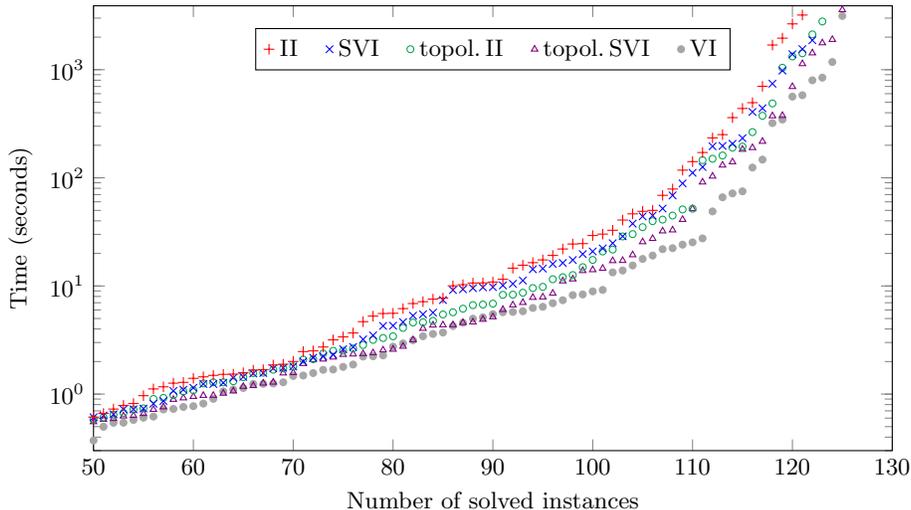

\cref{fig:experiments:time} depicts the model checking times for $\SVI$ (x-axis) and $\II$ (y-axis).
For better readability,  the benchmarks are divided into four plots with different scales.
Triangles ($\textcolor{blue}{\blacktriangle}$) and circles ($\textcolor{red}{\bullet}$) indicate MC and MDP benchmarks, respectively.
Similarly, \cref{fig:experiments:iters} shows the  required iterations of the approaches.
We observe that $\SVI$ converged faster and required fewer iterations for almost all MCs and MDPs.
$\SVI$ performed particularly well on the challenging instances where  many iterations are required.
Similar observations were made when comparing the topological variants of $\SVI$ and $\II$.
Both approaches were still competitive if no a priori bounds are given to $\SVI$.
More details are given in \cref{app:experiments}.

\cref{fig:experiments:all} indicates the model checking times of $\SVI$ and $\II$ as well as their topological variants.
For reference, we also consider standard (unsound) value iteration $(\VI)$.
The $x$-axis depicts the number of instances that have been solved by the corresponding approach within the time limit indicated on the $y$-axis. 
Hence, a point $(x,y)$ means that  for $x$ instances the model checking time was less or equal than $y$.
We observe that the topological variant of $\SVI$ yielded the best run times among all sound approaches and even competes with (unsound) $\VI$.  
  \section{Conclusion}
In this paper we presented  a sound variant of the value iteration algorithm which safely approximates reachability probabilities and expected rewards in MCs and MDPs.
Experiments on a large set of benchmarks indicate that our approach is a reasonable alternative to the recently proposed interval iteration algorithm. 
 \iftoggle{tr}{}{\pagebreak}%
  \bibliographystyle{splncs}
	\bibliography{literature}
\iftoggle{tr}{%
\clearpage
\appendix  
  \section{Correctness of Sound Gauss-Seidel Value Iteration}
\label{app:gs}
We provide additional details on the correctness of the Gauss-Seidel variant of our approach as presented in \cref{sec:dtmc:algorithm}.
Let $\dtmcDef$ be a DTMC with absorbing set of goal states $G$ and partition $\States = \StatesZero \cupdot G \cupdot \StatesMaybe$.
Further let ${\prec} \in \States \times \States$ be an arbitrary strict total order on the states.

We consider the functions 
$f_\prec \colon \RR^{|\States|} \to \RR^{|\States|}$ and
$h_\prec \colon \RR^{|\States|} \to \RR^{|\States|}$ with 
$\vectoraccessSet{f_\prec}{\StatesZero} = 0$, $\vectoraccessSet{f_\prec}{G} = 1$,  
$\vectoraccessSet{h_\prec}{\States \setminus
 \StatesMaybe} = 0$, 
\begin{align}
\vectoraccess{f_\prec(x)}{s} &=
\sum_{s' \prec s} \probP(s, s') \cdot \vectoraccess{f_\prec(x)}{s'} +
\sum_{s' \not\prec s} \probP(s, s') \cdot \vectoraccess{x}{s'} \text{, and }\\
\vectoraccess{h_\prec(y)}{s} &=
\sum_{s' \prec s} \probP(s, s') \cdot \vectoraccess{h_\prec(y)}{s'} +
\sum_{s' \not\prec s} \probP(s, s') \cdot \vectoraccess{y}{s'}.
\end{align}
The Gauss-Seidel variant of \cref{alg:dtmc} is obtained by replacing $f$ with $f_\prec$ and $h$ with $h_\prec$.
The modified algorithm still yields a sound $\precision$-approximation of the reachability probability $\reachpr$.

To show this, observe that for $k \ge 0$,  $x,y \in \RR^{|S|}$ with $\vectoraccessSet{x}{G} = 1$, $\vectoraccessSet{x}{\States \setminus G} = 0$, $\vectoraccessSet{y}{\StatesMaybe} = 1$, $\vectoraccessSet{y}{\States \setminus \StatesMaybe} = 0$ we have
\[
\vectoraccess{f_\prec^k(x)}{s} = \Prob_s(\eventually_\kappa^{\le k} G)
\text{ and }
\vectoraccess{h_\prec^k(y)}{s} = \Prob_s(\globally_\kappa^{\le k} \StatesMaybe).
\]
Here, $\eventually_\kappa^{\le k} G$ and $\globally_\kappa^{\le k} \StatesMaybe$ are defined as follows.
Let $\kappa \colon \States \times \States \to \{0,1\}$ with $\kappa(s, s') = 0$ if $s' \prec s$ and $\kappa(s,s') = 1$ if $s \prec s'$. We set
\begin{align*}
	\eventually_\kappa^{\le k} G &= \left\{ s_0 s_1 \dots \in \Paths[\dtmc] \mid  s_j \in G \text{ for some } j\ge 0 \text{ with } \sum_{i=0}^{j-1} \kappa(s_i, s_{i+1}) \le k  \right\}\\
	\globally_\kappa^{\le k} \StatesMaybe  &= \left\{ s_0 s_1 \dots \in \Paths[\dtmc] \mid  s_j \in \StatesMaybe \text{ for all }  j \text{ with }  \sum_{i=0}^{j-1} \kappa(s_i, s_{i+1}) \le k  \right\}.
\end{align*}
Intuitively, the paths in $\eventually_\kappa^{\le k} G$ reach $G$ within $k$ steps, where only steps from $s$ to $s'$ with $s \prec s'$ are counted.

The correctness of the Gauss-Seidel variant of \cref{alg:dtmc} follows from the following theorem which is analogue to \cref{thm:dtmc:prob} and can be shown in a similar way.
\begin{theorem}
	\label{thm:dtmc:gs}
	For DTMC $\dtmc$ let $G$, $\StatesMaybe$, $\kappa$ be as above and $k\ge0$  such that $\Prob_s(\globally_\kappa^{\le k} \StatesMaybe ) < 1$  for all $s \in \StatesMaybe$. 
	We have
	\begin{align*}
	&\Prob(\eventually_\kappa^{\le k} G) + \Prob( \globally_\kappa^{\le k} \StatesMaybe) \cdot \min_{s \in \StatesMaybe} \frac{	\Prob_s(\eventually_\kappa^{\le k} G) }{1 - \Prob_s(\globally_\kappa^{\le k} S_{?})} \\
	\le\, & \reachpr\\
	\le\, &	\Prob(\eventually_\kappa^{\le k} G) + \Prob( \globally_\kappa^{\le k} \StatesMaybe) \cdot \max_{s \in \StatesMaybe} \frac{	\Prob_s(\eventually_\kappa^{\le k} G) }{1 - \Prob_s(\globally_\kappa^{\le k} S_{?})}.
	\end{align*}
\end{theorem}

%

\section{Additional Experimental Results}\label{app:experiments}
We provide additional results of our experiments in order to substantiate our claims from \cref{sec:experiments}.
\begin{figure}[H]
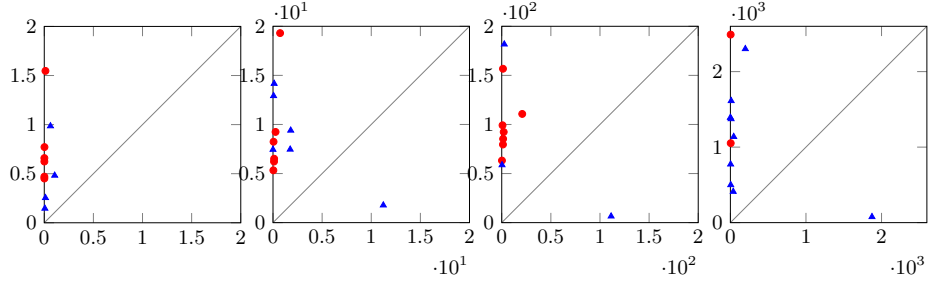

	\centering
	\adjustbox{max width=\linewidth}{
			\setplotlimits{0}{2}{-1}{20}{-2}{200}{-3}{2600}\scatterplot{SVI-time}{PRISM-time}
	}
	\caption{Comparison of model checking times for our implementation of $\SVI$ (x-axis) and the implementation of $\II$ in \prism (y-axis).}
			\label{fig:experiments:prism}
\end{figure}
\begin{figure}[h]
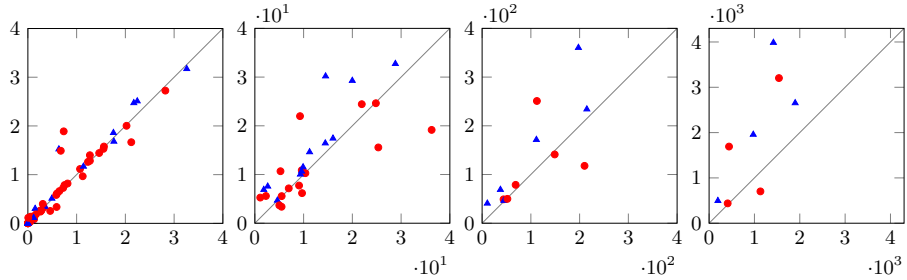
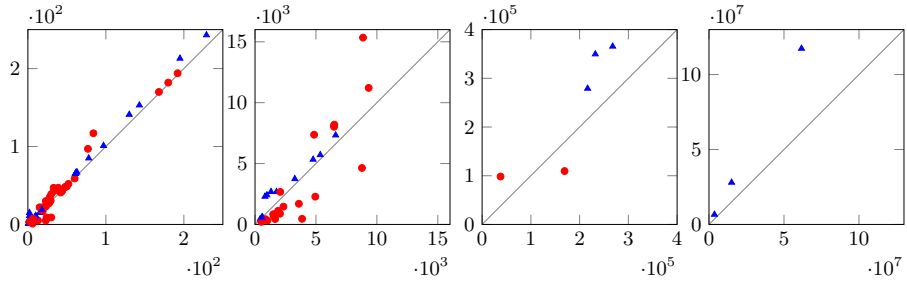

	\centering
	\adjustbox{max width=\linewidth}{
		\subfigure[Mmodel checking times (in seconds).]{%
			\label{fig:experiments:nb:time}
			\setplotlimits{0}{4}{-1}{40}{-2}{400}{-3}{4300}\scatterplot{nbSVI-time}{II-time}
	}}
	\adjustbox{max width=\linewidth}{
		\subfigure[Required iterations.]{%
			\label{fig:experiments:nb:iters}
			\setplotlimits{-2}{250}{-3}{16000}{-5}{400000}{-7}{130000000}\scatterplot{nbSVI-iters}{II-iters}	}}
	\caption{Comparison of $\SVI$ without computation of a priori bounds (x-axis) and $\II$ (y-axis).}
	\label{fig:experiments:nb}
\end{figure}

In \cref{fig:experiments:prism} a comparison of $\SVI$ with the implementation of $\II$ in \prism is given. We consider $\prism$ 4.4 available on its website \url{www.prismmodelchecker.org}.
Notice that we only consider a subset of our benchmark instances that are supported by both tools. 
In particular, \prism currently does not support Markov automata or interval iteration for multi-objective queries.

In \cref{fig:experiments:nb}  we compare $\II$ with a variant of $\SVI$ for which no initial bounds $u_0, \ell_0$ were computed.

The topological variants of $\SVI$ and  $\II$  are compared in \cref{fig:experiments:topo}.

\begin{figure}[h]
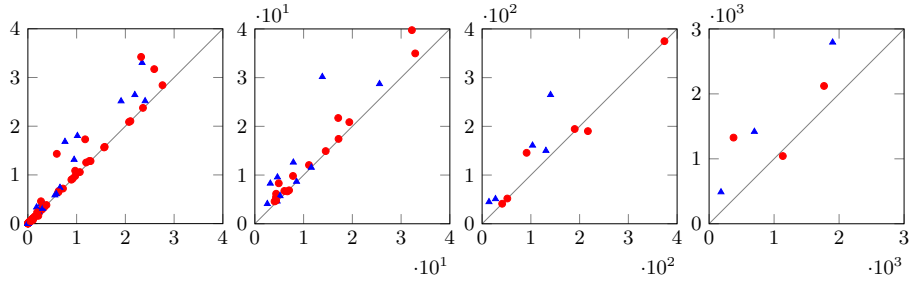
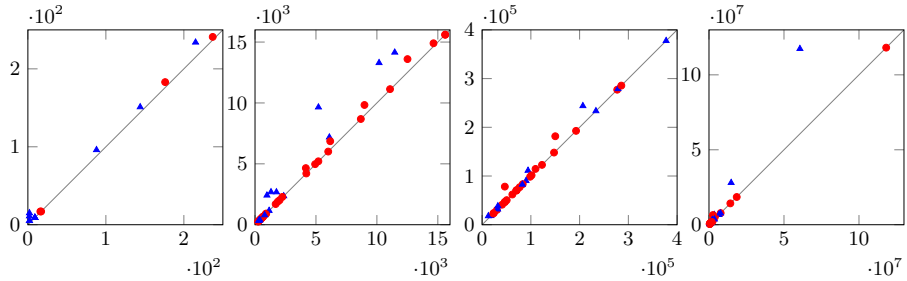

	\centering
	\adjustbox{max width=\linewidth}{
		\subfigure[Model checking times (in seconds).]{%
			\label{fig:experiments:topo:time}
			\setplotlimits{0}{4}{-1}{40}{-2}{400}{-3}{3000}\scatterplot{topoSVI-time}{topoII-time}
	}}
	\adjustbox{max width=\linewidth}{
		\subfigure[Required iterations.]{%
			\label{fig:experiments:topo:iters}
			\setplotlimits{-2}{250}{-3}{16000}{-5}{400000}{-7}{130000000}\scatterplot{topoSVI-iters}{topoII-iters}	}}
	\caption{Comparison of topological variants of $\SVI$ (x-axis) and $\II$ (y-axis).}
	\label{fig:experiments:topo}
\end{figure}

}{}%
\end{document}